\documentclass{article}
\usepackage{ijcai23}

\usepackage{times}
\usepackage{soul}
\usepackage{url}
\usepackage[hidelinks]{hyperref}
\usepackage[utf8]{inputenc}
\usepackage[small]{caption}
\usepackage{graphicx}
\usepackage{amsmath}
\usepackage{amsthm}
\usepackage{booktabs}
\usepackage{algorithm}
\usepackage{algorithmic}
\usepackage[switch]{lineno}

\usepackage{amssymb}
\usepackage{subfig}
\usepackage{amsmath}
\usepackage{makecell}
\usepackage{tabularx}
\usepackage{threeparttable}
\usepackage{multirow}
\usepackage{booktabs}
\usepackage{marvosym}
\usepackage{bm}
\usepackage{mwe}
\newcommand{\eg}{\emph{e.g.,}\xspace}

\newcommand{\ie}{\emph{i.e.,}\xspace}

\newcommand{\etal}{\emph{et al.}\xspace}

\let\oldhat\hat
\renewcommand{\vec}[1]{\mathbf{#1}}
\renewcommand{\hat}[1]{\oldhat{\mathbf{#1}}}

\usepackage{bibentry}

\title{Towards Hierarchical Policy Learning for Conversational Recommendation with Hypergraph-based Reinforcement Learning}

\author{
Sen Zhao$^{1,2}$\and
Wei Wei\thanks{Corresponding author}$^{1,2}$\and
Yifan Liu$^{1,2}$\and
Ziyang Wang$^{1,2}$\and
Wendi Li$^{1,2}$\and
Xian-Ling Mao$^3$\and
Shuai Zhu$^{4}$\and
Minghui Yang$^{4}$\and
Zujie Wen$^{4}$\\
\affiliations
$^1$Cognitive Computing and Intelligent Information Processing (CCIIP) Laboratory, School of Computer Science and Technology, Huazhong University of Science and Technology\\
$^2$Joint Laboratory of HUST and Pingan Property \& Casualty Research (HPL)\\
$^3$School of Computer Science and Technology, Beijing Institute of Technology\\
$^4$Ant Group
\emails
\{senzhao, weiw, yifaan, ziyang1997, wendili0822\}@hust.edu.cn,
maoxl@bit.edu.cn,
\{zs261988, minghui.ymh,  zujie.wzj\}@antgroup.com}

\begin{document}

\maketitle

\begin{abstract}
    Conversational recommendation systems (CRS) aim to timely and proactively acquire user dynamic preferred attributes through conversations for item recommendation.
     In each turn of CRS, there naturally have two decision-making processes with different roles that influence each other: 1) \textbf{director}, which is to select the follow-up \textbf{option} (\ie ask or recommend) that is 
    more effective for reducing the action space and acquiring user preferences; and 2) \textbf{actor}, which is to accordingly choose \textbf{primitive actions} (\ie asked attribute or recommended item) that satisfy user preferences and give feedback to estimate the effectiveness of the director's option.
    However, existing methods heavily rely on a unified decision-making module or heuristic rules, while neglecting to distinguish the roles of different decision procedures, as well as the mutual influences between them.
    To address this, we propose a novel \textbf{D}irector-\textbf{A}ctor \textbf{H}ierarchical  \textbf{C}onversational \textbf{R}ecommender (DAHCR), where the director selects the most effective option, followed by the actor accordingly choosing primitive actions that satisfy user preferences. Specifically, we develop a dynamic hypergraph to model user preferences and introduce an intrinsic motivation to train from weak supervision over the director. Finally, to alleviate the bad effect of model bias on the mutual influence between the director and actor, we model the director's option by sampling from a categorical distribution. Extensive experiments demonstrate that DAHCR outperforms state-of-the-art methods.
\end{abstract}

\section{Introduction}
\begin{figure}[ht]
    \centering\includegraphics[width=0.48\textwidth]{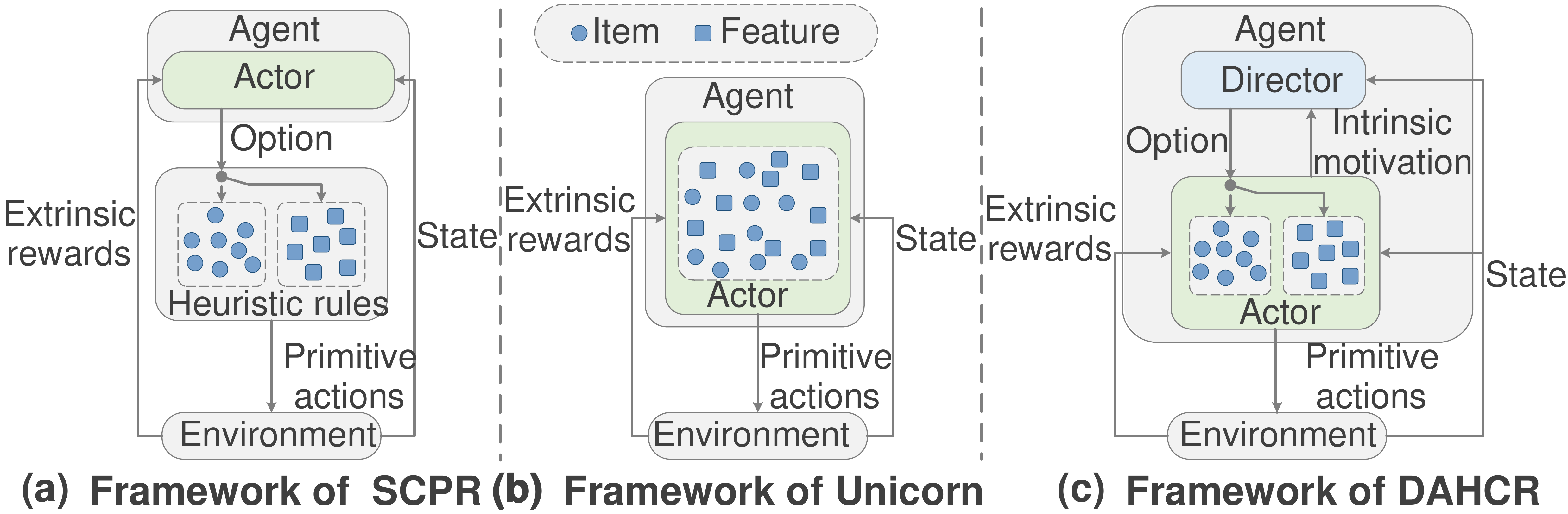}
    \caption{Illustration of policy learning frameworks for CRS, including a framework of the outsourcing strategy (SCPR), a unified framework (Unicorn), and our proposed Director-Actor framework.}
    \label{fig:motivation}
  \end{figure}
Conversational recommendation systems (CRS) aim to dynamically learn user preferences by iteratively interacting with the user.
Existing works have explored various settings of CRS from the perspective of either dialogue systems \cite{li2018towards} or recommendation systems \cite{lei2018sequicity}. This work focuses on the setting of multi-round conversational recommendation (MCR) \cite{sun2018conversational}, which aims to recommend the target item to the user by iteratively asking attributes and recommending items in the limited turns.

For each turn in CRS, the system naturally includes two essential decision-make procedures, when to recommend (\ie ask or recommend), and what to talk about (\ie the specific attribute/items). Early works \cite{lei2020estimation,sun2018conversational} develop policy learning for a subset of decision procedures and outsource the other procedures to heuristic rules (SCPR as illustrated in Figure~\ref{fig:motivation} (a)).
These works isolate strategies for different decisions and make policy learning hard to converge due to their lack of mutual influence during training. To solve this problem, Deng \etal \shortcite{deng2021unified} and Zhang \etal \shortcite{zhang2022multiple} develop unified policy learning frameworks (Unicorn as illustrated in Figure~\ref{fig:motivation} (b)) which unify the aforementioned two separated decision-make processes as a selection from the action space consisting of items and attributes. Despite effectiveness, the unified strategy brings out issues to be solved:  
(i) The unified strategy complicates the action selection of the CRS strategy by enlarging the action space and introducing data bias into the action space due to the imbalance in the number of items and attributes. As illustrated in Figure \ref{fig:motivation} (b), the action space is enlarged with all the items and attributes, and the strategy will prefer to select items when the number of items is larger.
(ii) The unified strategy ignores the different roles of the two decision procedures, leading to the sub-optimal CRS strategy. 

In the real scenario of CRS, the two decision procedures have different roles which are mutually influenced. As illustrated in Figure \ref{fig:motivation} (c), the decision procedure of when to recommend works as a \textbf{director}, which should select the option (\ie ask or recommend) that is more effective for reducing the action space and acquiring user preferences (\eg avoid recommending when the user's preference is not certain enough) to guide the latter procedure. The latter procedure works as an \textbf{actor}, which should accordingly choose the primitive action (\ie the specific attribute/items) that satisfies the user’s preference and gives feedback to evaluate the effectiveness of the director’s option.
The director's option limits the actor's action space to either attributes or items, which reduces the action space and avoids the data bias introduced by the imbalance in the number of items and attributes.  

There remain three challenges in modeling these two roles and their mutual influence.
The first challenge is weak supervision. The extrinsic rewards from the environment in each turn estimate the user's preference for the actor's primitive actions, but fail to estimate the effectiveness of the director's option, which is weakly supervised by the final-turn result (\ie success or failure). The second challenge is user preference modeling. In the scenario of CRS, the user likes/dislikes items since they satisfy some attributes, which is a three-order relation (\ie user-attribute-item). To specify the attributes that motivate the user to like/dislike the item, we should model user preferences with such high-order relations. The third challenge is the bad effect of model bias \cite{battaglia2018relational,tarvainen2017mean} on the mutual influence between director and actor. Specifically, the director’s bias may lead to bad options that will filter out more efficient actions for the actor. And the actor's bias can result in false feedback that will disturb the convergence of the director.

To overcome the aforementioned challenges, we propose a \textbf{D}irector-\textbf{A}ctor \textbf{H}ierarchical  \textbf{C}onversational \textbf{R}ecommender (DAHCR) with the director to select the option (\ie ask or recommend) that is more effective for reducing the action space and acquiring user preferences, followed by the actor accordingly choosing primitive actions (\ie specific items/attributes) that satisfy user preferences. To train from weak supervision over the director's option effectiveness, we develop and introduce an intrinsic motivation (\ie the actor's feedback) \cite{chentanez2004intrinsically} into our Director-Actor framework to estimate the effectiveness of director's options. Furthermore, to model user preferences, we develop a dynamic hypergraph \cite{feng2019hypergraph,jiang2019dynamic} with each high-order relation (\ie user-attribute-item) specifying an attribute that motivates the user to like/dislike the item. Finally, to alleviate bad effects of the model bias in the mutual influence, we model the director's option by sampling from a categorical distribution with Gumbel-softmax \cite{pei2022transformer}. Extensive experiments on real-world datasets show that our method outperforms the state-of-the-art methods.

In a nutshell, this work makes the following contributions:
\begin{itemize}
    \item We emphasize the different roles in two decision procedures for CRS, and the mutual influence between them.

    \item We propose a novel Director-Actor Hierarchical conversational recommender with intrinsic motivation to train from weak supervision and a dynamic hypergraph to learn user preferences from high-order relations. To alleviate the bad effect of model bias on the mutual influence between director and actor, DAHCR models the director's options by sampling from a categorical distribution with Gumbel-softmax.

    \item We conduct extensive experiments on two benchmark datasets, and DAHCR  effectively improves the performance of conversational recommendation.
\end{itemize}
\section{ Related Work}
Different from traditional recommendation systems \cite{zhao2022multi,wang2022user,wei2023recommendation} that predict the user's preference based on his/her historical behaviors, conversational recommendation systems (CRS) \cite{priyogi2019preference,xie2021comparison,zhou2020improving,zhao2023multi} aim to communicate with the user and recommend items based on the attributes explicitly asked during the conversation. Various efforts have been conducted to explore the challenges in CRS which can mainly be categorized into two tasks: dialogue-biased CRS studies the dialogue understanding and generation \cite{chen2019towards,kang2020recommendation,liu2020towards}, and recommendation-biased CRS explores the strategy to consult and recommend  \cite{christakopoulou2016towards,christakopoulou2018q,sun2018conversational,lei2020estimation}. This work focuses on the multi-round recommendation-biased CRS (MCR) \cite{lei2020estimation} which focuses on the setting where the MCR aims to recommend the target item to the user by iteratively asking attributes and recommending items in limited turns.

For each turn in CRS, the system naturally includes two make-decision procedures, when to recommend (\ie ask or recommend), and what to talk about (\ie which attribute/item to inquire/recommend).  Early works for the MCR improve the strategies of when and what attributes to ask, while the decision of which item to recommend is made by external heuristic rules. EAR \cite{lei2020estimation} utilizes latent vectors to capture the current state of MCR, and employs policy gradient to improve the strategy of deciding when to ask questions about attributes and which attribute to ask. To reduce the action space in policy learning, SCPR \cite{lei2020interactive} improves the strategy to only decide whether to ask or recommend and develops external path reasoning methods to decide which attribute to ask or which item to recommend. These works, however, isolate strategies for different problems and make the policy learning of these strategies hard to converge. To solve this problem, Unicorn \cite{deng2021unified} unifies the two decision procedures as the selection from the candidate action space consisting of items and attributes. Specifically, Unicorn proposes a graph-based Markov Decision Process (MDP) environment to choose actions from the candidate action space. MCMIPL \cite{zhang2022multiple} further considers the user's multiple interests in the unified strategy and develops a multi-interest policy learning module. Despite effectiveness, these works ignore the variant roles of different decision procedures, which may lead to sub-optimal CRS strategies.
\section{The Proposed Model}
We first introduce the problem definition of multi-turn conversational recommendation (MCR). Next, we introduce the framework and the model of our proposed  Director-Actor Hierarchical  Conversational Recommender (DAHCR).
\begin{figure*}[!ht]
    \centering\includegraphics[width=0.9\textwidth]{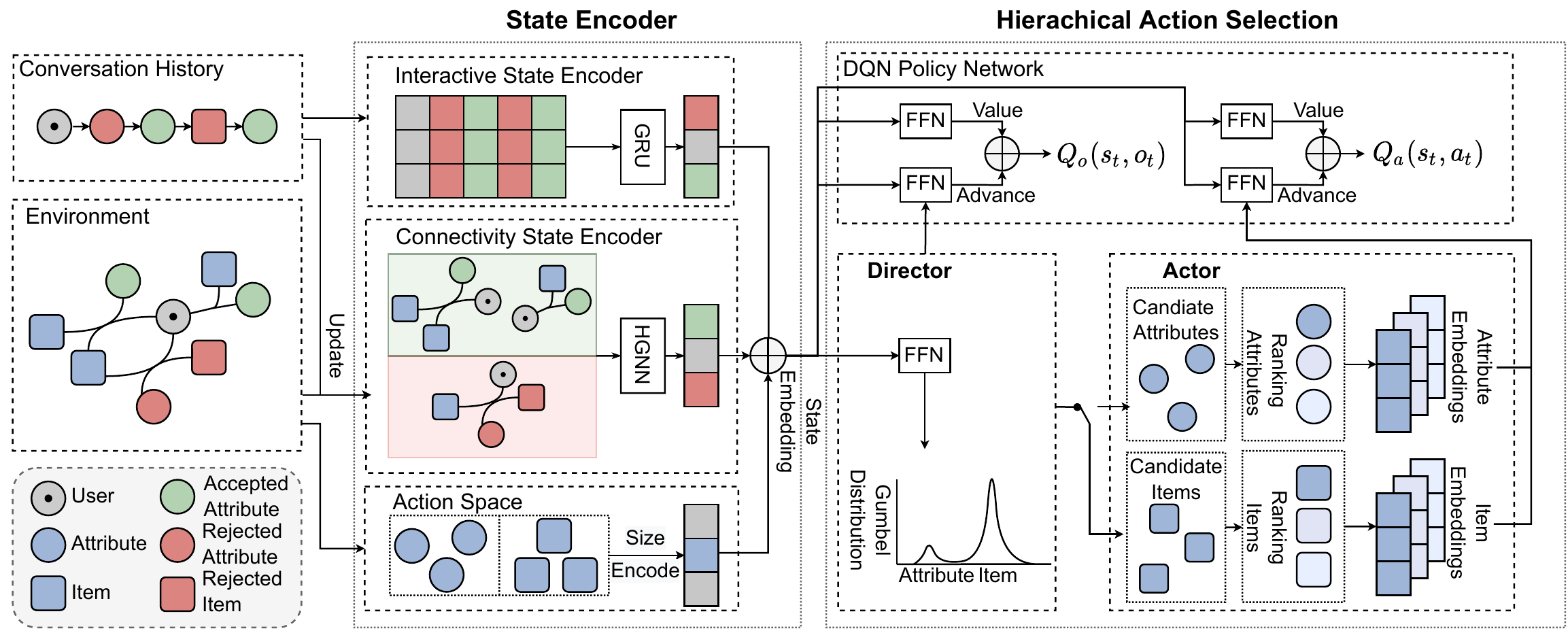}
    \caption{The overview of Director-Actor Hierarchical Conversational Recommender (Best view in color).}
    \label{fig:model}
  \end{figure*}
  
\subsection{Problem Formulation}
\label{sec:formu}
In this section, we formulate the problem of multi-turn conversational recommendation (MCR), which aims to recommend the target item to the user by asking attributes and recommending items in the limited turns of the conversation.

Specifically, let $V=\{v_1, v_2,\cdots, v_M\}$ denotes the item set. For each item $v$, there exists an attribute set $P_v$ associated with the item. At the beginning of each conversation, a user $u$ initializes the conversation session with a target item $v^*$ and an attribute that belongs to the target item $p_0 \in P_{v^*}$. The candidate item set $V_{cand}$ is formed with the items associated with $p_0$ and the candidate attribute set $P_{cand}$ is constructed by the attributes associated with the items in the candidate item set $V_{cand}$. Then at each turn $t$, MCR can either ask the user an attribute $p_t \in P_{cand}$ or recommend a certain number of items (\eg the top ten items) $V_t \subseteq V_{cand}$ to the user. According to the target item  $v^*$ and its associated attributes $P_{v^*}$, the user will choose to accept or reject the proposal of MCR. Based on the user's feedback, MCR will update the candidate attribute set $P_{cand}$ and the candidate item set $V_{cand}$. The conversation will continue until the max turn $T$ and the recommendation is successful if the target item $v^*$ is recommended within $T$.

\subsection{DAHCR Framework}
As illustrated in Figure \ref{fig:motivation} (c), we propose the Director-Actor hierarchical conversational recommendation policy Learning, a novel framework for MCR. At each time step $t$, the director chooses an option $o_t \in \mathcal{O}$, and the actor chooses the primitive action $a_t \in \mathcal{A}_{t|o_t}$ accordingly. Consequently, the state is updated to $s_{t+1}$ with the transition $T(s_{t+1}|s_t, o_t, a_t)$. The user's feedback $f_t \in \{acc, rej\}$, the extrinsic reward $r^a_t\in \mathcal{R}^a$, and intrinsic motivation $r^o_t\in \mathcal{R}^o$ are given according to $s_t$, $o_t$ and $a_t$.  Specifically, the main components of DAHCR $\langle\mathcal{S}, \mathcal{O}, \mathcal{A}, T, \mathcal{R}^o, \mathcal{R}^a\rangle$ are defined as:

\paragraph{\textbf{State} $\mathcal{S}$.}
The current state contains three components, including the interactive history $\mathcal{I}^{t}$, the related nodes $\mathcal{N}^{t}$, and the hypergraph $\mathcal{G}^{t}$ among the user and related nodes:

\begin{equation}
    s_t=[\mathcal{I}^{t}, \mathcal{N}^{t},  \mathcal{G}^{t}],
    \label{eq:state}
    \end{equation}
where $\mathcal{I}^t=[(a^1_j, f_j) | j =1, 2\cdots t-1]$. The related nodes $\mathcal{N}^t=\{u\}\cup P^t_{acc}\cup P^t_{rej}\cup V^t_{rej}\cup V^t_{cand}$ contains the user $u$, the accepted attributes $P^t_{acc}$, the rejected attributes $ P^t_{rej}$, the rejected items $V^t_{rej}$, and the candidate items $V^t_{cand}$. 

\paragraph{\textbf{Options} $\mathcal{O}$.} Based on the state $s_t$ of the current turn, the director should choose an option $o_t \in \mathcal{O}$, where $\mathcal{O}=\{ask, rec\}$ denotes whether to ask or recommend.

\paragraph{\textbf{Primitive actions} $\mathcal{A}$.} Based on the director's option $o_t$ and the state $s_t$, the actor selects the primitive action $a_t \in \mathcal{A}_{t|o_t}$, where $\mathcal{A}_{t|ask}=P^t_{cand}$ is the candidate attributes to ask and $\mathcal{A}_{t|rec}=V^t_{cand}$ means the candidate items to recommend. 

\paragraph{\textbf{Transitions} $T$.} As previous works \cite{deng2021unified}, the state $s_{t+1}$ updates based on  the user's response: 
\begin{equation}
    \begin{cases}
    P^{t+1}_{acc}=P^t_{acc} \cup a_t, & \text { if } o_t = ask, f_t=acc \\
    P^{t+1}_{rej}=P^t_{rej} \cup a_t, & \text { if } a_t = ask, f_t=rej \\
    V^{t+1}_{rej}=V^t_{rej} \cup a_t, & \text { if } a_t = rec, f_t=rej\end{cases},
    \label{eq:stateupd}
    \end{equation}
 \begin{equation}
    V^t_{cand}=V_{P^t_{acc}}\setminus V^t_{rej}, ~P^t_{cand}=P_{V^t_{cand}}\setminus (P^t_{acc} \cup P^t_{rej}),
    \label{eq:stateupd2}
    \end{equation}
 where $V_{P^t_{acc}}$ denotes the items that satisfy all the accepted attributes and $P_{V^t_{cand}}$ denotes all the attributes that belong to the candidate items. Follows Lei \etal \shortcite{lei2020interactive}, items that have rejected attributes are not eliminated from $V^t_{cand}$.

\paragraph{\textbf{Extrinsic rewards} $\mathcal{R}^a$.} Extrinsic rewards are special signals to guide the agent to select user-preferred actions. Five kinds of rewards are designed \cite{deng2021unified}: (1) $r^a_{acc|rec}$, a strongly positive reward for successful recommendation; (2) $r^a_{rej|rec}$, a slightly negative reward for rejected recommendation; (3) $r^a_{acc|ask}$, a slightly positive reward when the asked attribute is accepted; (4) $r^a_{rej|ask}$, a slightly negative reward when the asked attribute is rejected; (5) $r^a_{quit}$, a strong negative reward when the maximum turn reaches.

\paragraph{\textbf{Intrinsic motivation} $\mathcal{R}^o$.} The intrinsic motivation is passed from the actor to the director to estimate the effectiveness of the director's option.  Since recommending is an inefficient action when the user's preference is not certain enough (\ie the target item is outside the top ten in the actor's ranking list of items), we assign a positive reward $r^o_{+}$ to the option of $ask$ and a negative reward $r^o_{-}$ to the option of $rec$ in this situation. Inversely, when the user's preference is certain, $r^o_{+}$ and $r^o_{-}$ are assigned to the option of $rec$ and $ask$, respectively.

\subsection{DAHCR Policy Learning}
\subsubsection{State Encoder} 
The interactive history  $\mathcal{I}^t=[(o_j, f_j) | j \in \{1, 2\cdots t-1\}]$ between DAHCR and the user contains the director's historical options $o_j$ and the user's feedback of accepting or rejecting $f_j \in \{acc, rej\}$. With the embedding of the interactive history $\{\mathbf{X}^1_h, \mathbf{X}^2_h, \cdots,  \mathbf{X}^{t-1}_h \}$, the interactive state $\vec{s}^t_h$ is obtained with GRU networks \cite{cho2014properties}:
\begin{equation}
    \vec{s}^t_h=GRU(\vec{s}^{t-1}_h, \mathbf{X}^{t-1}_h).
\label{eq:his_state}
\end{equation}
The interactive state $\vec{s}^t_h$ encodes the historical interaction between the agent and the user, and is excepted to guide DAHCR to learn CRS strategy to decide whether the user's preference is certain enough (\eg the recommended attribute is accepted for several turns) for recommending items.
 
 To learn the user's preference for the specific attributes and items, we build a dynamic hypergraph $\mathcal{G}^{(t)}_u=(\mathcal{N}^{(t)}, \mathcal{H}^{(t)}, \mathbf{A}^{(t)})$, including: (1) the set of related nodes $\mathcal{N}^t=\{u\}\cup P^t_{acc}\cup P^t_{rej}\cup V^t_{rej}\cup V^t_{cand}$; (2) a hyperedge set $\mathcal{H}^{(t)}$, whose element $h\in \mathcal{H}^{(t)}$ denotes a hyperedge between the user, an attribute and items. In our case, for each attribute $p \in \mathcal{N}^{(t)}$, we define a hyperedge $h_p$ corresponding to the attribute $p$; (3) a $|\mathcal{N}^{(t)}|\times |\mathcal{H}^{(t)}|$ adjacent matrix $\mathbf{A}^{(t)}$ which denotes the weighted edge between each node and hyperedge, with entries denoted as:
\begin{equation}
    A_{i, j}^{(t)}= \begin{cases}
    1, & \text { if } n_{i}=u, p_{h_j} \in \mathcal{P}^{(t)}_{acc} \\
    -1, & \text { if } n_{i}=u, p_{h_j} \in \mathcal{P}^{(t)}_{rej} \\
    \frac{1}{|\mathcal{V}^{(t)}_{h_j}|}, & \text { if } n_{i}\in \mathcal{V}^{(t)}_{h_j} \\
    1, & \text { if } n_{i} = p_{h_j} \\ 
    0, & \text { otherwise }\end{cases},
    \label{eq:hyperedge}
    \end{equation}
where $p_{h_j}$ denotes the attribute corresponding to the hyperedge $h_j$, and $\mathcal{V}^{(t)}_{h_j}$ indicates items that satisfy the attribute $p_{h_j}$.

To take advantage of the connectivity information from the dynamic hypergraph, we employ hypergraph neural networks \cite{xia2022hypergraph} to refine the node representation with structure and connectivity information. Firstly, we aggregate information propagated from nodes to related hyperedges:
\begin{equation}
    \boldsymbol{H}=\boldsymbol{D}_{h}^{-1} \boldsymbol{A}^{T} \boldsymbol{E} \boldsymbol{W}_n,
    \label{eq:hyperagg1}
    \end{equation}
where $\mathbf{E} \in \mathbb{R}^{|\mathcal{N}^{(t)}|\times d}$ denotes the initial embedding of related nodes $\mathcal{N}^{(t)}$, $\boldsymbol{W}_n \in \mathbb{R}^{d \times d}$ is the weight matrix, and $\boldsymbol{D}_{h}$ is the diagonal matrix denoting the degree of the hyperedges, which is defined as the number of nodes connected by hyperedges.

During the conversation, the hyperedges are successively generated when the user accepts or rejects the asked attribute. Moreover, the higher-level interactions between different hyperedges are also important in learning user preferences. To model the sequential information and hyperedge-wise feature interactions,  higher-level hypergraph layers further pass messages through the interactions between hyperedges as:
\begin{equation}
    \boldsymbol{H}^l_{f}=\text{MHSA}_f(\boldsymbol{H}^{l-1}_{f}, \boldsymbol{H}^{l-1}_{f}, \boldsymbol{H}^{l-1}_{f}),
    \label{eq:hyperagg2}
    \end{equation}
where $f \in\{acc, rej\}$, and $\text{MHSA}(\cdot)$ indicates the multi-head self attention \cite{vaswani2017attention}. Finally, we aggregate the information from the hyperedges to refine the nodes' representations $\mathbf{\Gamma}_l$ and then obtain the connectivity state $s^t_g$:
\begin{equation}
    s^t_g=\sum_{l}\mathbf{\Gamma}_l(u),\mathbf{\Gamma}_l=\text{ReLU}(\mathbf{A} \cdot \boldsymbol{H}^l).
    \label{eq:hyperstate}
    \end{equation}

With the interactive state $s^t_h$ and the connectivity state $s^t_g$, the final state is obtained by:
\begin{equation}
\begin{split}
    s^t= s^t_h \oplus s^t_g \oplus s^t_{len}, \\
\end{split}
\label{eq:state}
\end{equation}
where $n_i$ denotes the node in the hypergraph, $s^t_{len}$ encodes the size of the candidate item and attribute set by dividing the length $|V_{cand}|$ and  $|P_{cand}|$  into ten-digit binary features \cite{lei2020estimation}, since it is also an important basis for deciding to ask or recommend (\eg the recommendation is easier to be successful when the candidate items' size is small).

\subsubsection{Hierarchical Action Selection Strategy}
After obtaining the state encoding the interactive history and the user's preference, we design a novel dueling Q-network to conduct policy learning under the hierarchical structure. Following the basic assumption that the delayed rewards are discounted by a factor of $\gamma$, we define the Q-value $Q_o(s_t,o_t)$ and $Q_a(s_t,a_t)$ as the expected reward for the director's option $o_t$ and the actor's action $a_t$ based on the state $s_t$:
\begin{equation}
\begin{split}
    Q_o(s_t,o_t)=f^o_{\theta_V}(s_t)+ f^o_{\theta_A}(s_t, o_t),
\end{split}
\label{eq:strategy1}
\end{equation}
\begin{equation}
\begin{split}
    Q_a(s_t,a_t|o_t)=P^*(s_t,o_t)(f^a_{\theta_V}(s_t)+ f^a_{\theta_A}(s_t, a_t)),
\end{split}
\label{eq:strategy2}
\end{equation}
where the value function $f^o_{\theta_V}(\cdot)$ $f^a_{\theta_V}(\cdot)$, and the advantage function $f^o_{\theta_A}(\cdot)$ $f^a_{\theta_A}(\cdot)$ are four separate multi-layer perceptions (MLP). $P^*(s_t,o_t)$ controls the action space of the actor's actions by masking the Q-value of actions $a_t$ according to the director's option $o_t$. To realize a differentiable discrete sample of the director's option and alleviate the bad effect of model bias on the mutual influence between the director and actor, we model $o_t$ by sampling from a categorical distribution with Gumbel-softmax as:
\begin{equation}
\begin{split}
    P^*(s_t,o_t)=\frac{exp((log(P(s_t,o_t))+\epsilon)/\tau)}{\sum_{o \in \mathcal{O}}exp((log(P(s_t,o))+\epsilon)/\tau)},
\end{split}
\label{eq:strategy3}
\end{equation}
where $\epsilon=-log(-log(x))$ and $x$ is sampled from $Uniform(0, 1)$. The temperature parameter $\tau$ controls the bias and variance of the likelihood distribution. When $\tau$ is larger, the likelihood is smoother with more variance and less bias. $P(s_t,o_t)$ is calculated with the softmax function as:
\begin{equation}
\begin{split}
    P(s_t,o_t)=\frac{exp(Q_o(s_t,o_t))}{\sum_{o \in \mathcal{O}} Q_o(s_t,o)},
\end{split}
\label{eq:strategy4}
\end{equation}
where $\mathcal{O}=\{ask, rec\}$ denotes the space for the director's option.  Gumbel-Softmax is used to alleviate the bias when lacking policy learning in the mutual influence between Director and Actor: (1)Biases (e.g., bad options of Director) caused by Director may filter out of efficient actions for the Actor. Hence, we consider sampling options from a Gumbel distribution, which may pave the way to explore the right action for Actor; (2)Bias (e.g., false feedback) caused by Actor may affect the convergence of Director, and thus to alleviate the issue, we consider to optimize the Gumbel distribution, rather than the deterministic function of Director’s option.

The optimal Q-function with the maximum expected reward $Q^*_o(s_t,o_t)$ and $Q^*_a(s_t,a_t)$ for the director's option and the actor's primitive action are achieved by optimizing the hierarchical policy $\pi_o$ and $\pi_a$,  follows the Bellman function \cite{bellman1957role} as:

\begin{equation}
\begin{split}
    Q^*_o(s_t,o_t)=\mathbb{E}_{s_{t+1}}[r^o_t+\gamma \underset{o_{t+1}\in \mathcal{O}}{max}Q^*_o(s_{t+1}, o_{t+1}|o_t)],
\end{split}
\label{eq:strategy4}
\end{equation}

\begin{equation}
\begin{split}
    Q^*_a(s_t,a_t)=\mathbb{E}_{s_{t+1}}[r^a_t+\gamma \underset{a_{t+1}\in \mathcal{A}_{t+1|o_{t+1}}}{max}Q^*_a(s_{t+1},a_{t+1}|a_t)],
\end{split}
\label{eq:strategy5}
\end{equation}
where $\mathcal{A}_{t+1|o_{t+1}}$ denotes the action space of primitive actions according to the director's option (\ie $\mathcal{A}_{t+1|ask}$ denotes candidate attributes and $\mathcal{A}_{t+1|rec}$ denotes candidate items). 

\subsubsection{Model Training} 
For each turn, the agent gets the intrinsic motivation $r^o_t$ to the director's option, the extrinsic reward $r^a_t$ to the actor's primitive action, and the candidate actions space $\mathcal{A}_{t+1}$ is updated according to the user's feedback. We define a replay buffer $\mathcal{D}$ following Deng \etal \shortcite{deng2021unified}, which stores the experience $(s_t, o_t, a_t, r^o_t, r^a_t, s_{t+1}, \mathcal{A}_{t+1})$. For the training procedure, we sample mini-batch from the buffer and optimize the model with loss function as follows:
\begin{equation}
\begin{split}
    \mathcal{L}_1(\theta_Q)=\mathbb{E}_{(s_t, o_t, r^o_t, s_{t+1})\sim\mathcal{D}}[(y^o_t-Q_o(s_t, o_t;\theta_Q))^2],
\end{split}
\label{eq:optimize1}
\end{equation}

\begin{equation}
\begin{split}
    \mathcal{L}_2(\theta_Q)=\mathbb{E}_{(s_t, a_t, r^a_t, s_{t+1}, \mathcal{A}_{t+1} )\sim\mathcal{D}}[(y^a_t-Q_a(s_t, a_t;\theta_Q))^2].
\end{split}
\label{eq:optimize2}
\end{equation}
$\mathcal{L}_1$ and $\mathcal{L}_2$ are alternatively optimized to teach DAHCR efficiently interacts with the user and predict the user's preference, where $\theta_Q=\{\theta_V, \theta_A\}$, and $y^o_t$, $y^a_t$ are target values for the director's options and the actor's actions, which are based on the optimal value function as: 
\begin{equation}
\begin{split}
    y^o_t=r^o_t+\gamma \underset{o_{t+1}\in \mathcal{O}}{max}Q^*_o(s_{t+1},o_{t+1};\theta_Q),
\end{split}
\label{eq:optimize3}
\end{equation}
\begin{equation}
\begin{split}
    y^a_t=r^a_t+\gamma \underset{a_{t+1}\in \mathcal{A}_{t+1}}{max}Q^*_a(s_{t+2},a_{t+1};\theta_Q).
\end{split}
\label{eq:optimize4}
\end{equation}
To alleviate the problem of overestimation bias, we adopt the double Q-learning \cite{van2016deep} to employ target networks $Q^{'}_o$ and $Q^{'}_a$ as period copies from the online networks to train the network following previous works \cite{deng2021unified,lei2020interactive}. 
\begin{table}[!htbp]
    \centering
    \small
        \begin{tabular}{p{1.5cm} p{1.0cm}<{\raggedleft}p{1.0cm}<{\raggedleft}p{1.25cm}<{\raggedleft}p{1.25cm}<{\raggedleft}}  
    \toprule
    \textbf{Dataset}   & \textbf{LastFM} & \textbf{LastFM*} &\textbf{Yelp}&\textbf{Yelp*}\\
    \midrule
      Users&1,801&1,801&27,675&27,675\cr

    Items&7,432&7,432&{70,311}&{70,311}\cr
    Interactions&76,693&{76,693}&{1,368,606}&{1,368,606}\cr
    Attributes&33&8,438&29&590\cr
    \hline
    Entities&{9,266}&{17,671}&{98,605}&{98,576}\cr
    Relations&{4}&{4}&{3}&3\cr
    Triples&138,215 &{228,217}&2,884,567 &2,533,827\cr
    \bottomrule
    \end{tabular}
    \caption{Statistics of datasets. }
    \label{tab:data}
\end{table}

\section{ Experiments}
To fully demonstrate the superiority of our method, we conduct experiments to verify the following three research questions (RQ): 
\begin{itemize}
    \item \textbf{RQ1}: Compared with the state-of-the-art methods, does our framework achieves better performance?
    \item \textbf{RQ2}: What are the impacts of key components on performance? 
    \item \textbf{RQ3}: How do hyper-parameters settings (such as the layer number of hypergraph neural networks) affect our framework?
\end{itemize}

\subsection{Experiment Setting}
\subsubsection{Datasets}
For better comparison, we follow previous works to conduct experiments\footnote{https://github.com/Snnzhao/DAHCR} on LastFM, LastFM* for music artist recommendation and Yelp, Yelp* for the business recommendation. The statistics of datasets are illustrated in Table~\ref{tab:data}.
\begin{itemize}
    \item {\textbf{LastFM}}~\cite{lei2020estimation}: LastFM is designed to evaluate the binary question scenario for the music artist recommendation, where the user gives preference towards an attribute using yes or no. Following Lei \etal \shortcite{lei2020estimation}, we manually merge relevant attributes into 33 coarse-grained attributes.
    \item{\textbf{Yelp}}~\cite{lei2020estimation}: Yelp is designed for enumerated questions for the business recommendation, where the user can select multiple attributes under one category. 
    \item{\textbf{LastFM*} and \textbf{Yelp*}}~\cite{lei2020interactive}: Following Lei \etal \shortcite{lei2020interactive}, we construct the datasets with original attributes and pruning off the attributes with frequency $<$ 10, named LastFM* (containing 8438 attributes) and Yelp* (containing 590 attributes) separately. 
\end{itemize}
\begin{table*}[t]
    \centering
    \small
    \label{results}
    \begin{tabular}{p{1.8cm}<{\centering}p{0.8cm}<{\centering}p{0.6cm}<{\centering}p{0.8cm}<{\centering}p{0.01cm}p{0.8cm}<{\centering}p{0.6cm}<{\centering}p{0.8cm}<{\centering}p{0.01cm}p{0.8cm}<{\centering}p{0.6cm}<{\centering}p{0.8cm}<{\centering}p{0.01cm}p{0.8cm}<{\centering}p{0.6cm}<{\centering}p{0.8cm}<{\centering}}
    \toprule
    \multirow{2}{*}{\bfseries Models }&\multicolumn{3}{c}{\bfseries LastFM }&&\multicolumn{3}{c}{\bfseries LastFM* }&&\multicolumn{3}{c}{\bfseries Yelp }&&\multicolumn{3}{c}{\bfseries Yelp* }\\
    \cline{2-4}
    \cline{6-8}
    \cline{10-12}
    \cline{14-16}
    &SR@15&AT&hDCG&&SR@15&AT&hDCG&&SR@15&AT&hDCG&&SR@15&AT&hDCG\\
    \midrule

    Abs Greedy & 0.222 & 13.48 & 0.073 &       & 0.635 & 8.66  & 0.267 &       & 0.264 & 12.57 & 0.145 &       & 0.189 & 13.43 & 0.089 \\
    Max Entropy & 0.283 & 13.91 & 0.083 &       & 0.669 & 9.33  & 0.269 &       & 0.921 & 6.59  & 0.338 &       & 0.398 & 13.42 & 0.121 \\
    \midrule
    CRM   & 0.325 & 13.75 & 0.092 &       & 0.580 & 10.79 & 0.224 &       & 0.923 & 6.25  & 0.353 &       & 0.177 & 13.69 & 0.070 \\
    EAR   & 0.429 & 12.88 & 0.136 &       & 0.595 & 10.51 & 0.230 &       & 0.967 & 5.74  & 0.378 &       & 0.182 & 13.63 & 0.079 \\
    SCPR   & 0.465 & 12.86 & 0.139 &       & 0.709 & 8.43  & 0.317 &       & 0.973 & 5.67  & 0.382 &       & 0.489 & 12.62 & 0.159 \\
    \midrule
    UNICORN & 0.547 & 11.57 & 0.176 &       & 0.798 & 7.58  & 0.412 &       & \underline{0.985} & \underline{5.33}  & \underline{0.397} &       & 0.522 & 11.55 & 0.174 \\
    MCMIPL & \underline{0.633} & \underline{11.54} & \underline{0.191} &       & \underline{0.839} & \underline{6.89}  & \underline{0.412} &       & 0.981 & 5.65  & 0.387 &       & \underline{0.552}& \underline{11.31}& \underline{0.178} \\
    \midrule
    DAHCR  &  \textbf{0.712}$^\dagger$ & \textbf{10.83}$^\dagger$ & \textbf{0.213}$^\dagger$ &       & \textbf{0.925}$^\dagger$ & \textbf{6.31}$^\dagger$ & \textbf{0.431}$^\dagger$ &       &\textbf{0.992} & \textbf{5.16}$^\dagger$ & \textbf{0.400}$^\dagger$ &       & \textbf{0.626}$^\dagger$ & \textbf{11.02}$^\dagger$ &\textbf{0.192}$^\dagger$ \\
    \bottomrule
    \end{tabular}
    \caption{Experimental results.$^\dagger$ represents the improvement of DAHCR over all baselines is statistically significant with p-value $< 0.01$. hDCG indicates hDCG@(15, 10). SR@15 and hDCG are the higher the better, while AT is the lower the better.}
\label{tab:overall_result}
\end{table*}  
\subsubsection{Metrics}
Following previous studies \cite{lei2020estimation,lei2020interactive,deng2021unified}, we adopt three widely used metrics for conversational recommendation:
SR@t, AT, and hDCG.
Success rate (SR@t) is adopted to measure the cumulative ratio of successful recommendations by the turn t.
Average turns (AT) is used to evaluate the average number of turns for all sessions. 
hDCG@(T, K) is used to additionally evaluate the ranking performance of recommendations. 
Therefore, the higher SR@t and hDCG@(T, K) indicate better performance, while the lower AT means an overall higher efficiency.

\subsubsection{Implementation Details}
Following previous works \cite{deng2021unified,lei2020interactive}, we adopt TransE \cite{bordes2013translating} from OpenKE \cite{han2018openke} to pretrain the embedding of nodes in the constructed graph with the training set. The embedding size and the hidden size are set as 64 and 100. The temperature parameter $\tau$ is set to be 0.7 for Yelp* and 0.3 for the other three datasets. The layer number of hypergraph neural networks is selected from 1, 2, 3, and 4.
We set the intrinsic motivations as:  $r^o_{+}=1$,  $r^o_{-}=-1$.
The settings of the extrinsic rewards are the same as previous works \cite{lei2020estimation,lei2020interactive,deng2021unified}:
 $r^a_{rec\_{acc}}=1$,  $r^a_{rec\_{rej}}=-0.1$,  $r^a_{ask\_{acc}}=0.01$,  $r^a_{ask\_{rej}}=-0.1$, $r^a_{quit}=-0.3$.
In the training procedure, the size of the experience replay buffer is 50,000, and the batch size is 128. The learning rate and the $L_2$ norm regularization are set to be 1e-4 and 1e-6, with Adam optimizer.  The discount factor $\gamma$ is set as 0.999.
Our experiment is conducted on Nvidia RTX 3090 graphic cards equipped with an AMD r9-5900x CPU (32GB Memory).

\subsection{Baselines}
To demonstrate the effectiveness of the proposed DAHCR, we choose state-of-the-art methods for comparison. Specifically, we first choose two rule-based methods, three reinforcement learning-based methods that outsource part of decision procedures, and then two reinforcement learning (RL)-based method that unifies two decision procedures.
\begin{itemize}
    \item \textbf{Max Entropy \cite{lei2020estimation}.} This method employs a rule-based strategy to ask and recommend. It chooses to select an attribute with maximum entropy based on the current state, or recommends the top-ranked item with certain probabilities.
    \item \textbf{Abs Greedy \cite{christakopoulou2016towards}.} This method only makes the item-recommendation actions and updates the model based on the feedback. It keeps recommending items until the successful recommendation is made or the pre-defined round is reached. 
    \item \textbf{CRM \cite{sun2018conversational}.} A RL-based method that records the users' preferences into a belief tracker and learns the policy deciding when and which attributes to ask based on the belief tracker.
    \item \textbf{EAR \cite{lei2020estimation}.} This method proposes a three-stage solution to enhance the interaction between the conversational component and the recommendation component.
    \item \textbf{SCPR \cite{lei2020interactive}.} This method models CRS as an interactive path reasoning problem. It prunes irrelevant candidate attributes by traversing attribute vertices on the graph based on user feedback.
    \item \textbf{UNICORN \cite{deng2021unified}.} A RL-based approach that unifies the two decision strategies. It learns graph-enhanced state representations for RL via graph neural networks.
    \item \textbf{MCMIPL \cite{zhang2022multiple}.} A state-of-the-art approach to CRS that extends the unified conversational recommendation strategy with multi-interest representations of the user.
\end{itemize}

\subsection{Performance Comparison (RQ1)}
\subsubsection{Overall Performance} From the overall performance of all methods reported in Table \ref{tab:overall_result}, we make the following observations:

\begin{itemize}

\item \textbf{Our proposed DAHCR achieves the best performance.} DAHCR significantly outperforms all the baselines by achieving a higher success rate and hDCG, and less average turn on four datasets. The reason for such improvement can be attributed to the following aspects: 
i) With the hierarchical strategy, our proposed DAHCR can reduce the action space and avoid the data bias introduced by the imbalance in the number of items and attributes.
ii) The director and actor in DAHCR can well play their roles of choosing the effective option (\ie ask or recommend) and learning the user's preference over the candidate items and attributes.
iii) The intrinsic motivation can well deal with the problem of weak supervision on the director's option effectiveness.
iv) Modeling user preferences with dynamic hypergraph, DAHCR could specify the attributes that motivate the user to like/dislike the item with high-order relations.
v) Modeling the director's option by sampling from a categorical distribution can well alleviate the bad effect of model bias on the mutual influence between the director and the actor.

\item \textbf{Mutual influence of different decision procedures and the user's dynamic preference in the policy learning is important for CRS.}
From Table \ref{tab:overall_result}, DAHCR, MCMIPL and UNICORN surpasses CRM, EAR and SCPR in terms of three metrics over four datasets. There are two reasons for this performance: (i) CRM, EAR and SCPR isolate strategies for different decision procedures during the training process, which makes the conversational recommendation strategy hard to converge. This demonstrate the importance of the mutual influence between different decision procedures in the training process. (ii) Compared with other methods, DAHCR, MCMIPL and UNICORN use RL to update the user preferences on the attributes and items, which leads to better performance. This proves the importance of learning the user's dynamic preference for CRS.


\end{itemize}

\begin{figure}[t]
\centering
\begin{minipage}[t]{0.48\columnwidth}
\centering
\includegraphics[width=\columnwidth]{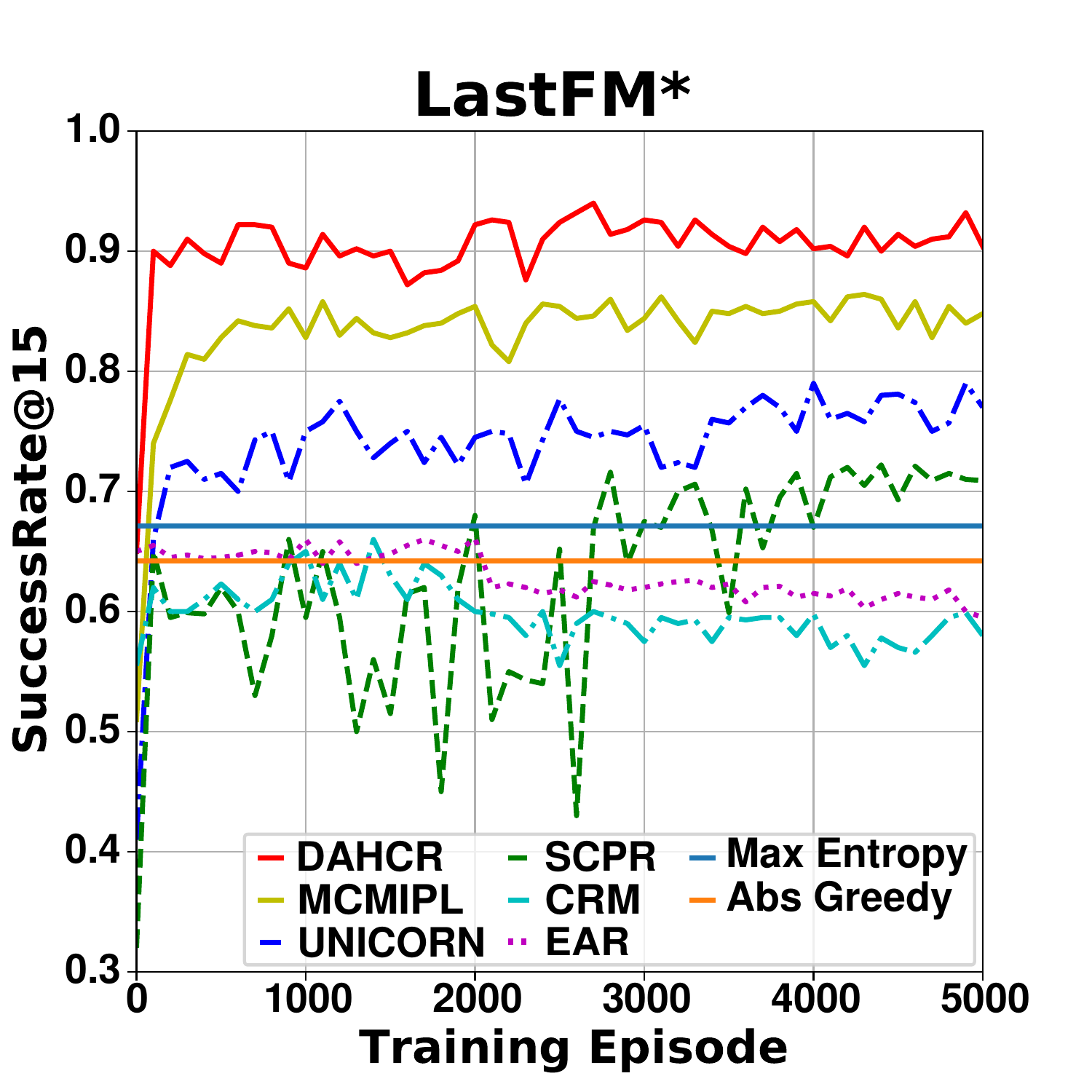}
\end{minipage}
\begin{minipage}[t]{0.48\columnwidth}
\centering
\includegraphics[width=\columnwidth]{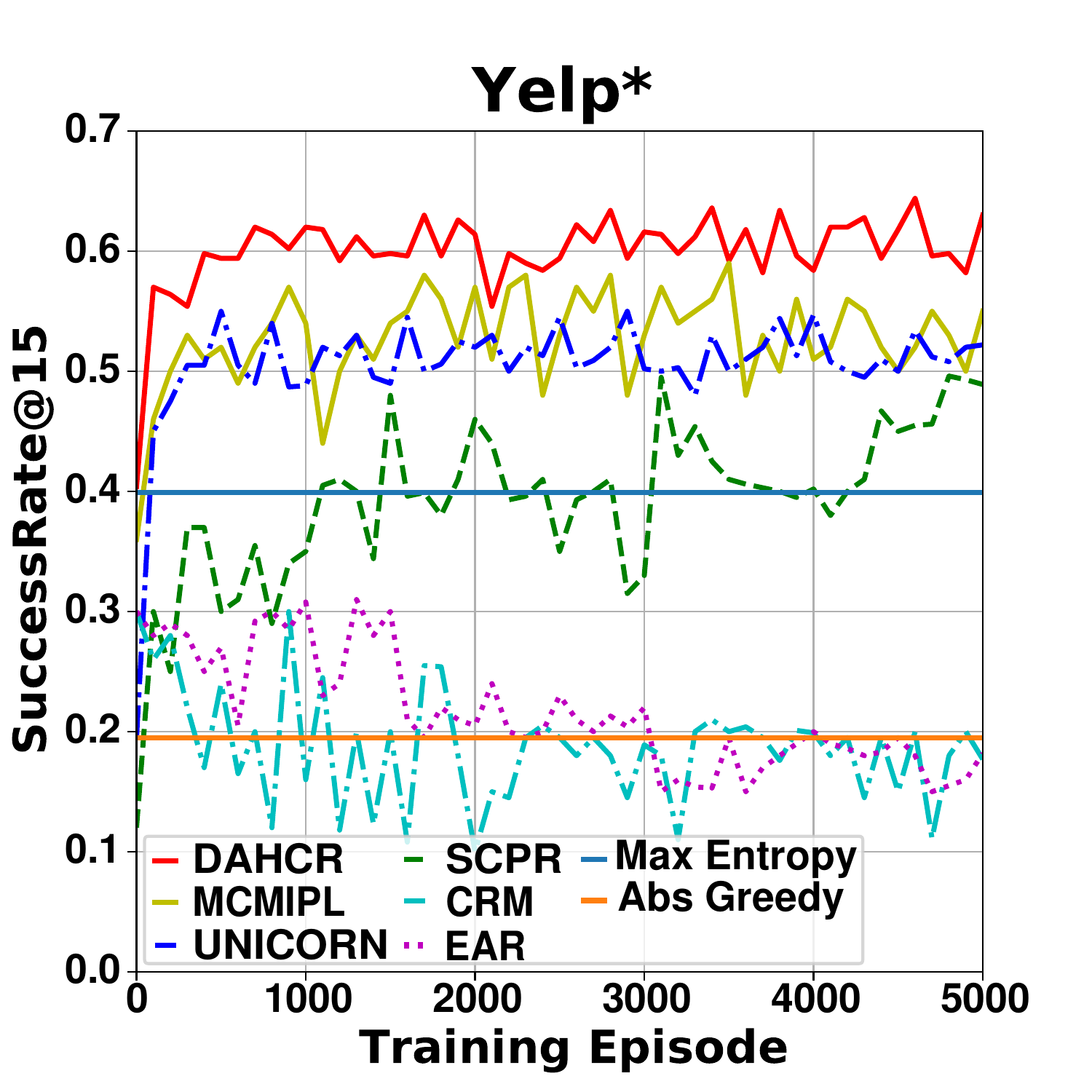}
\end{minipage}
\caption{Test performance at different training epochs.}
\label{fig:test_episodes_perform}
\end{figure}

\subsubsection{Training Efficiency} Figure~\ref{fig:test_episodes_perform} shows the performance curves of the different methods tested on LastFM* and Yelp*, respectively. The test performance curves for the unsupervised methods Max Entropy and Abs Greedy are shown as two horizontal lines for comparison.
It can be seen that DAHCR is far superior in converge speed and stability to all the baselines.
The reason is that the hierarchical conversational recommendation strategy of DAHCR can realize more effective action choice and better learn the user's preference.
In the $\text{Yelp}^*$, where the action candidate space is larger, the performance of EAR and CRM does not improve much or even gets worse as the training process iterates.
These results demonstrate the efficiency and effectiveness of the proposed DAHCR.
{
\begin{table}[t]
  \centering
  \normalsize 
    \resizebox{\columnwidth}{!}{
    \begin{tabular}{lccccccc}
    \toprule[1pt]
          & \multicolumn{3}{c}{LastFM*} &       & \multicolumn{3}{c}{Yelp*} \\
\cmidrule{2-4}\cmidrule{6-8}          & SR@15 & AT    & hDCG  &       & SR@15 & AT    & hDCG \\
    \midrule
    DAHCR w/o  & \textbf{0.925} & \textbf{6.31} & \textbf{0.431} &       & \textbf{0.626} & \textbf{11.02} & \textbf{0.192}  \\
    \midrule
    (a) Hie. & 0.842 & 6.88  & 0.415 && 0.551      &  11.35     & 0.176 \\
    (b) Hyper. & 0.909 & 6.54  & 0.428 &&  0.608&  11.13 & 0.189 \\
    (c) Gumbel. & 0.863 & 6.70  & 0.418 && 0.560      &  11.30  & 0.181 \\
    (d) Intrinsic. & 0.887 & 6.63  & 0.424 &  &  0.570   &   11.31    &  0.183       \\
    \bottomrule[1pt]
    \end{tabular}%
    }
    \caption{Results of the Ablation Study.}
  \label{tab:ablation}%
\end{table}%
}

\subsection{Study of DAHCR (RQ2\&RQ3)}
\begin{figure}[t]
\centering
    \includegraphics[width=0.45\textwidth]{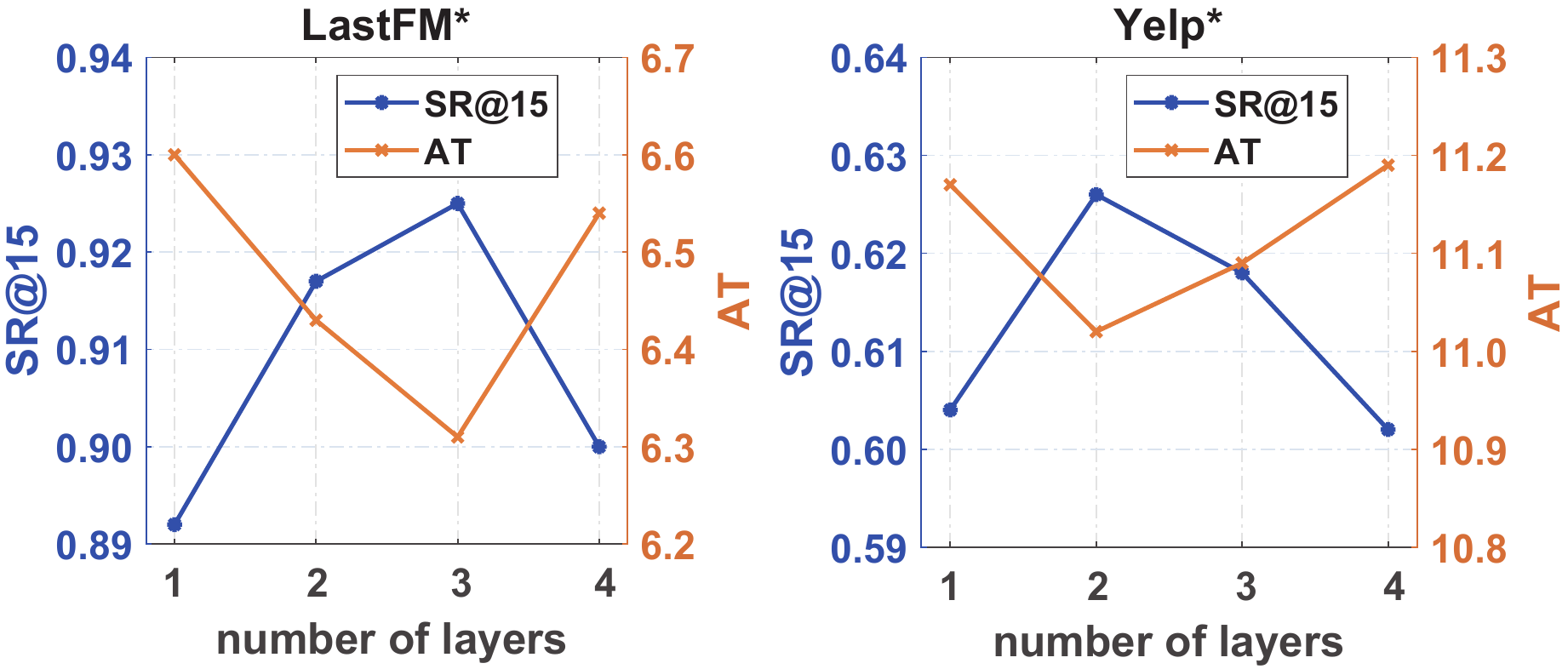}
    \caption{Impact of Layer Number(L)}
    \label{fig:layers}
    \vspace{-0.3cm}
  \end{figure}

Next we investigate the underlining mechanism of our DAHCR with four ablated models that remove the hierarchical framework, dynamic hypergraph, Gumbel-softmax, and intrinsic motivation, respectively. From the results in Table~\ref{tab:ablation}, we observe that:

\begin{itemize}
    \item The performance of DAHCR suffers a significant degradation when replacing the hierarchical framework with the unified framework. This demonstrates the importance of modeling the variant roles of different decision procedures in CRS.
    \item The performance of DAHCR drops when replacing the dynamic hypergraph neural networks with graph neural networks. We attribute this to the importance of high-order relations (\ie user-attribute-item) in modeling user preferences for CRS.
    \item The model performs worse when removing the Gumbel-softmax. This result suggests that alleviating the bad effect of model bias on the mutual influence between the director and the actor is necessary. Our method that models the director's option by sampling from the categorical distribution with Gumbel-softmax can reasonably deal with such an effect.
    \item The performance of the model drops when removing the intrinsic motivation, which indicates the necessity of intrinsic motivation to train DAHCR from weak supervision on the director's effectiveness.
\end{itemize} 
Figure~\ref{fig:layers} shows the experimental results by varying the layer number of hypergraph neural networks. Two-layer DAHCR performs better than one-layer DAHCR since it can capture high-order collaborative information in the dynamic hypergraph. But the performance of DAHCR does not always increase when the layer number increases. We attribute this to the noise that increases along with the hop of neighbors.

\section{Conclusion}
 In this work, we propose a Director-Actor Hierarchical Conversational Recommender (DAHCR) with the director to select the most effective option (\ie ask or recommend), followed by the actor accordingly choosing primitive actions that satisfy user preferences. The intrinsic motivation is designed for training from weak supervision on the director's effectiveness, a dynamic hypergraph is developed to learn user preferences from high-order relations, and Gumbel-softmax is employed to alleviate the bad effect of model bias on the mutual influence between director and actor. Experimental results on two real-world datasets show that the proposed DAHCR outperforms the state-of-the-art methods.
 \section*{Acknowledgments}
This work was supported in part by the National Natural Science Foundation of China under Grant No.62276110, Grant No.61772076, in part by CCF-AFSG Research Fund under Grant No.RF20210005, and in part by the fund of Joint Laboratory of HUST and Pingan Property \& Casualty Research (HPL). The authors would also like to thank the anonymous reviewers for their comments on improving the quality of this paper.

\bibliographystyle{named}
\bibliography{ijcai23}

\end{document}